\documentclass[aps,prb,amsmath,amssymb,twocolumn,showpacs]{revtex4}
\usepackage{bm}
\usepackage{graphicx}
\usepackage{color}  
\newcommand{\nix}[1]{}
\begin{document}
\title{Valley separation in graphene by polarized light}
\author{L.\,E.\,Golub and S.\,A.\,Tarasenko}
\affiliation{Ioffe Physical-Technical Institute of the RAS, 194021 St.\,Petersburg, Russia}
\author{M.\,V.\,Entin$^{(1)}$ and L.\,I.\,Magarill$^{(1,2)}$}
\affiliation{$^{(1)}$ Institute of Semiconductor Physics, Siberian
Branch of the RAS, 630090 Novosibirsk, Russia \\ $^{(2)}$ Novosibirsk State University, 630090 Novosibirsk, Russia}
\begin{abstract}
We show that the optical excitation of graphene with
polarized light leads to the pure valley current where carriers in
the valleys counterflow.
The current in each valley originates from asymmetry of
optical transitions and
electron scattering by impurities owing to the
warping of electron energy spectrum. 
The valley current has strong polarization dependence, its direction is opposite for normally incident beams of orthogonal linear polarizations.
In undoped graphene on a 
substrate with high susceptibility, 
electron-electron
scattering leads to an additional contribution to the valley
current that can dominate. 

\end{abstract}


\maketitle


Graphene, one-atom-thick layer of carbon with the honeycomb
crystal lattice, has been attracting rapidly growing attention due
to its unique electronic properties. Zero band gap
and zero effective electron and hole masses as well as high enough
mobility make it perspective
for fundamental and applied research.~\cite{Geim07,Neto09,Peres10}
The electron excitations in graphene are similar to massless Dirac
fermions with the cone points situated at the points $K$ and
$K'$ of the Brillouin zone. The interplay of two
equivalent valleys 
gives rise to new
transport and optical phenomena, which are absent in systems with
simple electron dispersion, and underlies the novel research field
called ``valleytronics''.~\cite{TarIvch05,Rycerz07} In multivalley
structures, one can independently control the carriers in
different valleys  and construct peculiar electron distribution
where particles in the valleys flow predominantly in different
directions.~\cite{Karch11}

Previous research of valley-dependent transport in graphene was
focused on the manipulation of charge carriers by static electric
field. It was demonstrated that the electric field may induce
valley-polarized current in a graphene point contact with zigzag edges,~\cite{Rycerz07}
graphene layer with broken inversion symmetry,~\cite{Xiao07}
bilayer graphene,~\cite{Abergel09} or if the structure is additionally
illuminated by circularly polarized radiation.~\cite{Oka09} 
It was also proposed in Ref.~[\onlinecite{Moskalenko09}] that valley currents can be induced in mesoscopic graphene rings by asymmetrical monocycle electromagnetic pulses.
Here, we show that the valley separation can be achieved  in a homogeneous graphene layer
by pure optical means.
We demonstrate that the interband excitation of graphene by
linearly polarized light leads to the electron
current in each valley,
which direction is determined by the light polarization. The partial
photocurrents $\bm{j}^{(\nu)}$ ($\nu=\pm $ for the valleys $K$ and $K'$, respectively)
in the ideal honeycomb structure are directed oppositely, so that the total
electric current $\bm{j}^{(+)} + \bm{j}^{(-)}$ vanishes.
We also briefly discuss optical and transport methods to reveal the pure valley current.

Phenomenologically, the emergence of the valley photocurrent is
related to the low point-group symmetry of individual valleys. Despite the fact that the
crystal lattice of flat graphene is centrosymmetric, the
valleys $K$ and $K'$ are described by the wave vector group
$D_{3h}$ lacking the space inversion, see Fig.~1. 
The group $D_{3h}$ allows for the
photocurrent induced by normally-incident linearly polarized
light. The polarization dependences of the current components in
the valley $K$ are given by
\begin{equation}\label{j_phen}
    j_x^{(+)}= \chi (e_x^2-e_y^2) I \:,
\quad j_y^{(+)}= - 2 \chi e_x e_y  I \:.
\end{equation}
Here, $\chi$ is a parameter, $e_x$ and $e_y$ are components of the
(real) light polarization unit vector $\bm{e}$, $I$ is the intensity of incident light, and the $x$ axis is chosen along the $\Gamma$-$K$ line,
Fig.~1a. The photocurrent in the valley $K'$ is obtained
from Eq.~(\ref{j_phen}) by the replacement $x \rightarrow -x$,
which gives $\bm{j}^{(-)}=-\bm{j}^{(+)}$.
We note that the absence of a total electric current at normal incidence of radiation is in agreement with the symmetry arguments allowing for a photocurrent in noncentrosymmetric systems only.
At oblique incidence of the radiation, a net current in graphene may
arise due to the photon drag effect.~\cite{Entin10,Karch10}

\begin{figure}[b]
\includegraphics[width=0.9\linewidth]{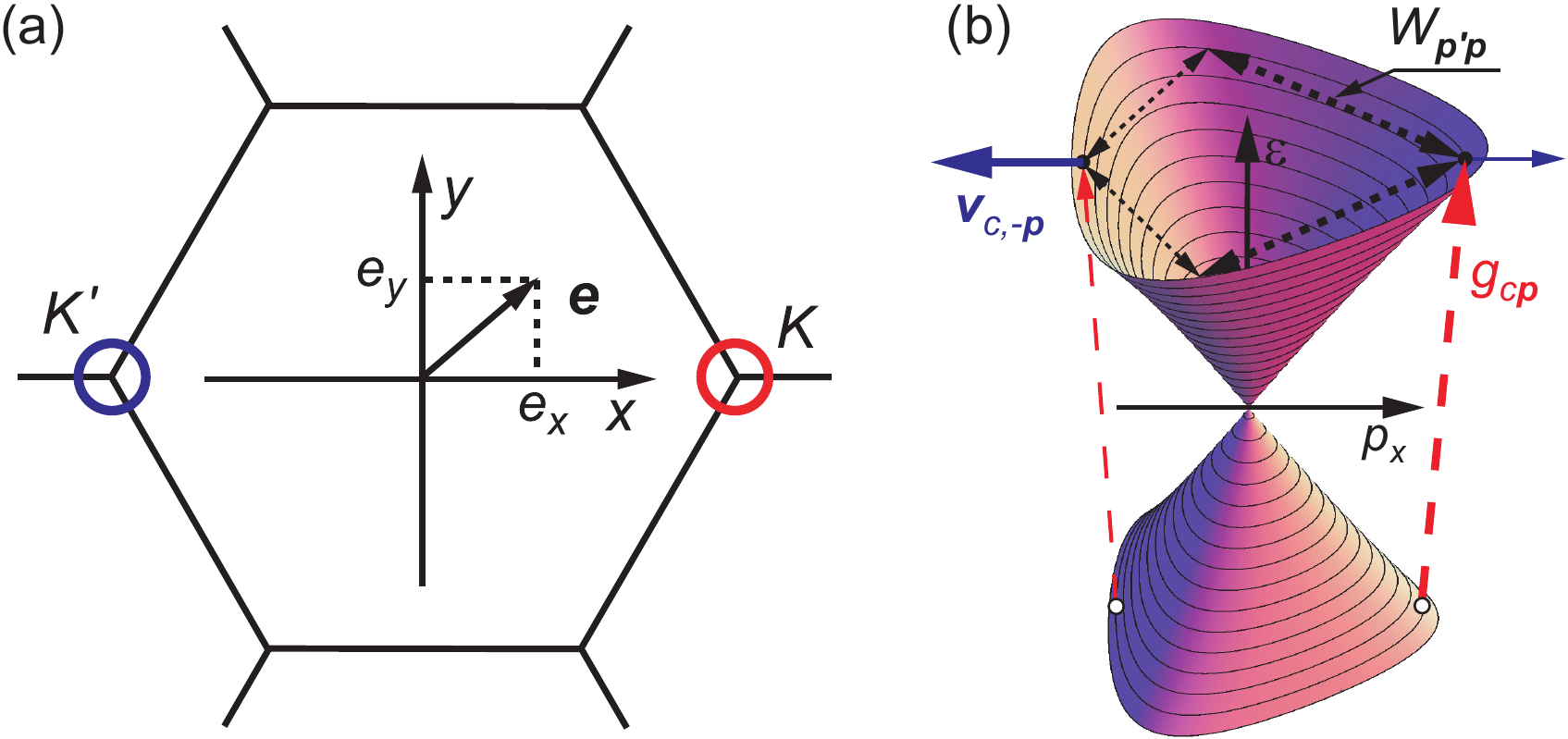}
\caption{(color online) (a) Brillouin zone of graphene.  The circles indicate neighborhood of the $K$ and $K'$ points where the electron states have trigonal symmetry allowing for the photocurrent. (b) 
Mechanisms of photocurrent formation in the $K$ valley. Solid, dashed, and dotted arrows of different thicknesses indicate anisotropy of velocity, optical generation, and scattering rate, respectively, in $\bm{p}$-space.}
\label{fig1}
\end{figure}
%


The microscopic model of pure valley current
generation is based on 
the trigonal warping of energy spectrum of carriers in the
valleys. The effective Hamiltonian describing electron and hole
states in the vicinity of the $K$ and $K'$ points has the form~\cite{McCann06}
\begin{equation}\label{Hamiltonian}
\hat{H}_{\bm p}^{(\nu)} = \left(
\begin{array}{cc}
0 & \Omega_{\bm p}^{(\nu)} \\
\Omega_{\bm p}^{(\nu)*} & 0
\end{array}
\right) \:.
\end{equation}
Here, $\bm p$ is the momentum counted from the valley center,
\begin{equation}\label{Omega}
\Omega_{\bm p}^{(\nu)} = \nu v_0 (p_x-{\rm i}p_y) - \mu (p_x + i
p_y)^2 \:,
\end{equation}
$v_0$ is the electron velocity, and $\mu$ is the parameter of warping
that reflects the trigonal symmetry of valleys ($D_{3h}$ wave vector group).
In the  framework of tight-binding model,
$\mu=v_0a/(4\sqrt{3}\hbar)$ with $a$ being the lattice constant.~\cite{Neto09}
We assume that the warping is 
small and, therefore, calculate the current to first order in
$\mu$. The energy spectrum of carriers
in the conduction ($c$) and valence ($v$) bands is given by
\begin{equation}\label{spectrum}
\varepsilon_{c{\bm p}}^{(\nu)} \approx  v_0 p  - \nu \mu p^2 \cos
3 \varphi_{\bm{p}}, 
\quad \varepsilon_{v{\bm p}}^{(\nu)}=-\varepsilon_{c{\bm p}}^{(\nu)} \:,
\end{equation}
where $\varphi_{\bm{p}}$ is the polar angle of the momentum $\bm{p}$.

Shown in Fig.~1b is the energy spectrum in the $K$ valley
with the warping being included. The inequality of
$\varepsilon_{c{\bm p}}^{(+)}$ and $\varepsilon_{c,-{\bm p}}^{(+)}$ (as well as $\varepsilon_{v{\bm p}}^{(+)}$ and $\varepsilon_{v,-{\bm p}}^{(+)}$) gives
rise to an electric current in the valley if electrons are optically excited from the valence to conduction band by linearly polarized light. In the valley $K'$, the warping of energy spectrum is opposite, Eq.~(\ref{spectrum}),
and the photocurrent direction is reversed.

In the framework of kinetic theory, the photocurrent densities in the valleys are given by
\begin{equation}\label{j_theory}
{\bm j}^{(\nu)} = 2 e \sum_{\bm p} \left( {\bm v}_{c\bm p}^{(\nu)} \,
f_{c\bm p}^{(\nu)} + {\bm v}_{v\bm p}^{(\nu)} \, f_{v\bm p}^{(\nu)} \right),
\end{equation}
where $e$ is the electron charge, 
the factor 2 accounts for spin degeneracy, ${\bm v}_{c,v}^{(\nu)} =\bm{\nabla}_{{\bm p}} \varepsilon_{c, v}^{(\nu)}$ are the velocities, and
$f_{c\bm p}^{(\nu)}$ and $f_{v\bm p}^{(\nu)}$ are the nonequilibrium corrections to the
distribution functions in the conduction and valence
bands linear in the light intensity; $f_{c\bm p}^{(+)} = f_{c,-\bm p}^{(-)}$ and $f_{v\bm p}^{(+)} = f_{v,-\bm p}^{(-)}$ due to space inversion symmetry.  We consider interband
optical transitions in undoped graphene at low temperature. Owing to electron-hole symmetry 
$f_{c\bm p}^{(\nu)}=-f_{v\bm p}^{(\nu)}$,
and the photocurrent Eq.~(\ref{j_theory})
assumes the form ${\bm j}^{(\nu)} = 4e \sum_{\bm p} {\bm v}_{c\bm p}^{(\nu)}
\, f_{c\bm p}^{(\nu)}$.

The steady-state correction to the distribution
function can be found from the kinetic equation
\begin{eqnarray}\label{kin_eq}
\sum_{{\bm p}'} \left( W_{{\bm p}{\bm p}'}^{(\nu)} \, f_{c{\bm
p}'}^{(\nu)} - W_{{\bm p}'{\bm p}}^{(\nu)} \, f_{c\bm p}^{(\nu)} \right) +
\mbox{St}^{({\rm ee})}+ g_{c\bm p}^{(\nu)}=0,
\end{eqnarray}
where $W_{{\bm p}{\bm p}'}^{(\nu)}$ is the rate of intravalley
electron scattering by static defects or impurities, the weak intervalley processes are neglected, 
$g_{c\bm p}^{(\nu)}$ is the optical generation rate, and
$\mbox{St}^{({\rm ee})}$ describes electron-electron collisions.

First, we consider the valley current in the presence of intensive electron scattering by impurities
and neglect electron-electron collisions. We focus on the photocurrent in the $K$ valley and omit
index $\nu=+$. In the Born approximation, the
rate of elastic electron scattering by impurities $W_{{\bm p}{\bm p}'}=W_{{\bm p}'{\bm p}}$ is given by 
\begin{equation}\label{W}
W_{{\bm p}{\bm p}'} = \frac{\pi}{2\hbar} \left| 1 +
\frac{\Omega_{\bm p}\Omega^*_{{\bm p}'}}{|\Omega_{\bm
p}\Omega_{{\bm p}'}|} \right|^2 {\cal K}(|{\bm p}'-{\bm p}|) \,
\delta(\varepsilon_{c\bm{p}}-\varepsilon_{c\bm{p}'}) \:,
\end{equation}
where ${\cal K}(q)$ is the Fourier component of the impurity
potential correlator.
The specific angular dependence of $W_{{\bm p}{\bm p}'}$ follows
from the Hamiltonian Eq.~(\ref{Hamiltonian}). The generation rate in the conduction band $g_{c\bm{p}}$ 
is determined by the interband matrix elements of the velocity operator $\bm{\nabla}_{\bm{p}} \hat{H}_{\bm p}$ and has the form
\begin{equation}\label{G} 
g_{c\bm p} = {2\pi\over\hbar} \left(eA\over c \right)^2
\left|{\rm Im} \left({\Omega_{\bm p}^* \over |\Omega_{\bm p}|} \,
{\bm e}\cdot {\bm \nabla}_{\bm p}\Omega_{\bm p}\right)\right|^2 \hspace{-1mm}
\delta(\hbar\omega - 2 \varepsilon_{c\bm{p}}) .
\end{equation}
Here $\omega$ is the light frequency, $A/2$ is the amplitude of
the vector potential of the electromagnetic wave related to the intensity of incident light
by $I=A^2\omega^2/(2\pi c t_0^2)$, and $t_0$ is the amplitude transmission coefficient,
$t_0 = 2/(n+1)$ for graphene on a substrate with the refractive index $n$.

As follows from Eqs.~(\ref{j_theory}) and~(\ref{kin_eq}), the
valley current arises owing to warping-induced asymmetry in the
electron velocity ${\bm v}_{c\bm p}$, the scattering rate
$W_{{\bm p}{\bm p}'}$, and the generation rate
$g_{c\bm{p}}$. Accordingly, to first order in $\mu$ one can
distinguish three contributions to the current Eq.~(\ref{j_phen}), $\chi=\chi^{({\rm vel})}+\chi^{({\rm gen})}+\chi^{({\rm sc})}$. The corresponding mechanisms of the current formation are sketched in Fig.~1b.

To calculate the valley current caused by the velocity correction
one neglects the warping in optical generation and scattering
rates. In this mechanism, the absorption of linearly polarized
light leads to the alignment of electron momenta described by the
second angular harmonic of the distribution function.\cite{OO_Milrin} Owing to the
$\mu$-linear correction to the velocity, such a distribution
of carriers in $\bm{p}$-space implies an electric current
Eq.~(\ref{j_phen}) with
\begin{equation}\label{vel}
\chi^{({\rm vel})}=  {5 e \mu \eta  \,
\tau_2(\varepsilon_\omega) t_0^2 \over 8 v_0} \:.
\end{equation}
Here, $\varepsilon_\omega=\hbar\omega/2$ is the kinetic energy of
photoelectrons, $\tau_n$ ($n=1,2\ldots$) are the relaxation times of the $n$th angular harmonics of the 
distribution function, $\tau_n^{-1} = \sum_{{\bm p}'} W_{{\bm
p}{\bm p}'} (1-\cos{n\theta})$, $\theta$ is the angle between $\bm{p}$ and $\bm{p}'$, and $\eta= \pi e^2/\hbar c$
is the absorbance (Ref.~[\onlinecite{Peres10}]) for normally-incident light.

Another contribution to the valley current comes from the
asymmetry of photoexcitation. Indeed, to first order in
$\mu$, the optical generation rate $g_{c\bm{p}}$ contains the
first angular harmonic, which gives rise to a photocurrent
\begin{equation}\label{gen}
\chi^{({\rm gen})}= -{e \mu \eta t_0^2 \over 8  v_0} \left[ 9
\tau_1(\varepsilon_\omega) +  \varepsilon_\omega {d
\tau_1(\varepsilon_\omega) \over d \varepsilon_\omega} \right] \:.
\end{equation}

The third mechanism of the current generation originates from the
asymmetry of electron scattering. The solution of kinetic
Eq.~(\ref{kin_eq}), $f_{c\bm p}$, contains the first angular
harmonic even if the warping is neglected in the optical
generation rate. Such a contribution to the valley current is
given by
\begin{eqnarray}
\label{sc}
\chi^{({\rm sc})} &=& {e \mu \eta  \, \tau_2 t_0^2 \over 8 v_0}  
\Biggl\{ 
20 - 6{\tau_1\over\tau_2} - 4{\tau_1\over\tau_3} \\
&+& {\varepsilon_\omega \over 2}
\biggl[
\left({9\over\tau_1}-{2\over\tau_2}\right) {d \tau_1 \over d \varepsilon_\omega}
+ \tau_1 { d\over d\varepsilon_\omega}\left({1\over\tau_2}+{1\over\tau_3}\right)
\biggr]
\Biggr\}, \nonumber
\end{eqnarray}
where the relaxation times are taken at the energy
$\varepsilon_{\omega}$.

Equations~(\ref{vel})-(\ref{sc}) demonstrate that both the magnitude and excitation spectrum of the pure valley current 
are determined by the mechanisms of scattering. For electron
scattering by unscreened Coulomb impurities in graphene, one obtains $\tau_1 \propto \varepsilon$, $\tau_2 =
3\tau_1$, and $\tau_3 = 5 \tau_1$. Such relations yield $\chi  \propto \omega$.
In the case of scattering by short-range static defects, one has
$\tau_1 \propto 1/\varepsilon$, $\tau_2=\tau_3 = \tau_1 /2$ and,
therefore, $\chi  \propto 1/\omega$.
Estimation shows that the valley currents in suspended graphene $j^{(\pm)} \sim 10^{-4}$~A/cm at the light intensity $I=1$~W/cm$^2$, $\tau_1=10^{-12}$~s, 
$\mu=3.6\times 10^{26}$~g$^{-1}$, and $v_0=10^8$~cm/s.


Now, we analyze the effect of electron-electron interaction on the pure valley current.
It is well known that, in systems with parabolic energy spectrum, the interparticle
collisions 
partially suppress the anisotropy of the distribution function. Therefore, one can expect that 
electron-electron scattering (between carriers from the same valley and, in particular, between carriers from different valleys) can only decrease the pure valley current. We demonstrate below that the interparticle collisions in graphene may give rise to an additional contribution to the valley photocurrent. 

Consider the collision of a photoelectron with the momentum ${\bm p}$ from the valley $\nu$ of the conduction band with an electron with the momentum ${\bm k}$ from the valley $\nu'$ of the valence band. After the collision, both electrons occur in the conduction band with the momenta ${\bm p}'$ and ${\bm k}'$, respectively, see the left inset in Fig.~2. 
Processes of other kinds are negligible in undoped graphene at low temperature and weak excitation level.   
The Coulomb interaction between carriers leads to the transfer of momentum $q \sim \omega/v_0 \ll \pi/a$, therefore, both electrons remain in the valleys they were before the collision although the valleys $\nu$ and $\nu'$ may be different.
The momentum and energy conservation laws read
\begin{equation}\label{Conservation}
{\bm p}+{\bm k}={\bm p}'+{\bm k}' \:, \quad \varepsilon_{c \bm p}^{(\nu)} + \varepsilon_{v \bm k}^{(\nu')} =
\varepsilon_{c \bm p'}^{(\nu)} + \varepsilon_{c \bm k'}^{(\nu')} \:.
\end{equation}
In the conic approximation, Eqs.~(\ref{Conservation}) imply that $k < p$,  $p'+k' < p$, and 
${\bm k}$ should be antiparallel while ${\bm p}'$ and ${\bm k}'$ parallel to $\bm p$.~\cite{Basko}
The possibility of the collisions is determined by corrections to the linear dispersion. 
Interaction-induced renormalization of the energy
spectrum makes it concave,~\cite{Mishchenko07,Elias11} which
forbids such Auger-like processes. On the contrary, the spectrum warping does
allow for the processes. In graphene on a substrate with high susceptibility,
the interaction effects are suppressed and the warping becomes prevailing.
An estimation shows that, for an electron with the energy $\varepsilon =0.3$~eV, this regime occurs at the effective dielectric constant $\epsilon^*>300$.
Besides, interaction effects can be suppressed by a metallic gate or increasing the temperature. Neglecting the interaction-induced renormalization of the energy
spectrum, we obtain from Eqs.~(\ref{Conservation}) to first order in $\mu$
\begin{eqnarray}\label{conserv2}
&& v_0 \left( p k \alpha^2 + p'k' \beta^2\right) = \;\;\; \\
&& - 2 \mu \nu (p-p')(p'+k')\left[p+p'+\nu\nu'(k-k')\right] \cos{3\varphi_{\bm p}}
 \:, \nonumber
\end{eqnarray} 
where $\alpha = \varphi_{\bm p}-\pi - \varphi_{\bm k} \ll 1$ and $\beta =\varphi_{\bm k'}-\varphi_{\bm p'}\ll 1$. 
Equation~(\ref{conserv2}) has solutions only for sectors of $\varphi_{\bm p}$ where the second line
is positive, i.e., for $\mu \nu \cos{3\varphi_{\bm p}}<0$.
It means that Auger-like processes for electrons with the momentum $\bm p$ are allowed or forbidden depending on the sign of $\cos{3\varphi_{\bm p}}$, see the right inset in Fig.~2.

\begin{figure}[t]
\includegraphics[width=0.9\linewidth]{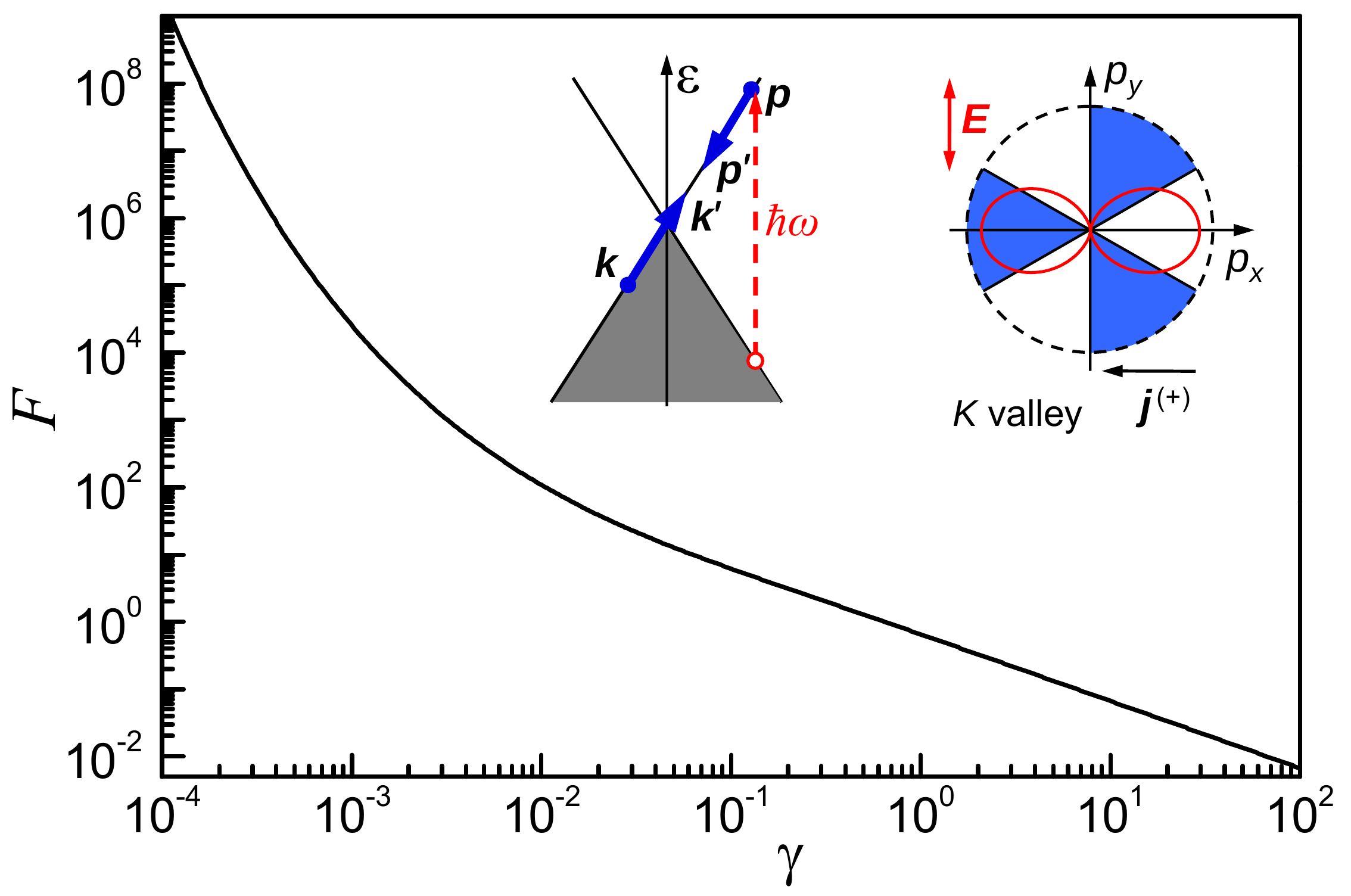}
\caption{(color online) Dependence $F(\gamma)$ that determines the magnitude of valley current caused by electron-electron scattering, Eq.~(\ref{F}). Left inset sketches scattering of electrons with the momenta $\bm{p}$ and $\bm{k}$ into the states with momenta $\bm{p}'$ and $\bm{k}'$. Right inset shows the momentum space in the $K$ valley. Sectors where the Auger-like electron relaxation is allowed are colored, ellipses depict the alignment of photoelectron momenta by linearly polarized light.}
\end{figure}
\label{fig2}

The angular dependence of the electron-electron collision rate in graphene gives rise to a pure valley current if the sample is illuminated by linearly polarized light. The mechanism of the current generation is sketched in the right inset in Fig.~2. The 
absorption of linearly polarized light leads to the alignment of electron momenta described by the second angular harmonic of the distribution function. The photoexcited electrons are scattered by resident electrons from the valence band and lose their energies.
Since the scattering rate is anisotropic, the energy relaxation leads to the formation of first harmonic of the distribution function and to an electric current if the momentum relaxation time depends on energy. 
In the case of weak electron-electron scattering, $\tau_{ee} \gg \tau_1$, where $\tau_{ee}$ is the electron-electron scattering time, the mechanism efficiency can be estimated as $\chi^{({\rm ee})} \sim e v_0 \eta \tau_1^2 /(\varepsilon_{\omega} \tau_{ee})$. 

To calculate the pure valley current caused by electron-electron scattering we solve the kinetic Eq.~(\ref{kin_eq}) with the linearized collision integral
\begin{eqnarray}\label{kin1}
&& \mbox{St}^{({\rm ee})} = \dfrac{4\pi}{\hbar} \sum_{{\bm k},{\bm k}',{\bm
p}',\nu'}|u_{{\bm p} -{\bm p}'}|^2 \, \delta_{{\bm p}+{\bm k},{\bm p}'+{\bm k}'} \\ 
\nonumber && \times
\Biggl\{ \left[
f_{c{\bm p}'}^{(\nu)} \, \delta(\varepsilon_{c \bm p}^{(\nu)} + \varepsilon_{c \bm k}^{(\nu')} - \varepsilon_{c {\bm p}'}^{(\nu)} - \varepsilon_{v {\bm k}'}^{(\nu')})   
- (\bm p \leftrightarrow \bm p')
\right]    
    \\ \nonumber
   && + \left[
    f_{c{\bm k}'}^{(\nu')} \, \delta(\varepsilon_{c \bm p}^{(\nu)} + \varepsilon_{c \bm k}^{(\nu')} - \varepsilon_{v {\bm p}'}^{(\nu)} - \varepsilon_{c {\bm k}'}^{(\nu')}) 
    + (\bm k \leftrightarrow \bm k')
    \right]\nonumber \Biggr\} \:,
\end{eqnarray}
where $u_{q}=2\pi \hbar e^2/(\epsilon^{*} q)$ is the Fourier component of the Coulomb potential
and the factor $4$ accounts for the spin degeneracy. In Eq.~(\ref{kin1}), we assume that the warping is small and take it into account only in the arguments of $\delta$-functions. The consequent simplification of ${\rm St}^{({\rm ee})}$ consists in summing over the almost collinear momenta.
Kinetic Eq.~(\ref{kin_eq}) with the simplified electron-impurity collision term $-f_{c \bm p}^{(\nu)}/\tau$ takes the form
\begin{eqnarray}\label{kin4}
g_{c \bm p} =  -
{\pi e^4 \over \hbar^3 v_0 \epsilon^{*2} p} \left[
\int\limits_p^\infty d k  \left(\sqrt{k} - \sqrt{p} \right)^2
\left( f_{c \bm k}^{(+)} + f_{c \bm k}^{(-)}\right)
\right. \,\,\,
\\
+
\left. \theta(-\nu\cos{3\varphi_{\bm p}}) \left(  \int\limits_p^\infty d k {\sqrt{k p} \over 2 }  f_{c \bm k}^{(\nu)} - {p^2\over 3 } f_{c \bm p}^{(\nu)} \right) 
\right] + {f_{c \bm p}^{(\nu)}\over \tau} \, ,
\nonumber
\end{eqnarray}
where $\varphi_{\bm k}=\varphi_{\bm p}$ and $\theta(x)$ is the Heaviside step function. 
In the case of 
$1/\tau(\varepsilon)=2 \pi^2 e^4 N_i/(\hbar  \epsilon^{*2} \varepsilon )$, which corresponds to the momentum relaxation time of electrons due to scattering by Coulomb impurities with the surface density $N_i$, 
Eq.~(\ref{kin4}) can be transformed into a differential
equation
\begin{eqnarray}\label{111}
    \gamma u (u^4+6\gamma)J''''(u) + \gamma(5u^4-16\gamma)J'''(u)+ \nonumber \\ 
    2u^2(u^4 + 12\gamma)J'(u) + 12u(u^4-2\gamma)J(u)=0
\end{eqnarray}
for the function
\[
J(u)= \dfrac{\varepsilon_{\omega}^2 \int\limits_0^{2\pi}
 f_{c \bm p}^{(+)} \, \theta(\cos3\varphi_{\bm p})\cos\varphi_{\bm p} d\varphi_{\bm p}}{ 2 \pi^2 v_0^2 \tau(\varepsilon_{\omega}) \sum_{\bm{p}} g_{c\bm{p}}  \, \theta(\cos3\varphi_{\bm p})\cos\varphi_{\bm p}} - \delta(u-1) \:.
\]
Here $u=\sqrt{v_0p/\varepsilon_\omega}$, $J'(u)=dJ(u)/du$, and 
the parameter $\gamma = \pi N_i \hbar^2 v_0^2/\varepsilon_\omega^2$ characterizes
the rate of electron scattering by impurities with respect to the electron-electron scattering rate.
$J(u)$ satisfies the boundary conditions:
$J(1)=J'(1)=0$, $J''(1)=2(1+12\gamma)/[\gamma(1+6\gamma)]$, and
$J'''(1)=-36/(1+6\gamma)^2$. Finally, the contribution to the valley current induced by electron-electron scattering is given by Eq.~(\ref{j_phen}) with
\begin{equation}\label{F}
	\chi^{({\rm ee})}= - \frac{e v_0  \eta \tau (\varepsilon_\omega) t_0^2}{4\pi \varepsilon_\omega} F(\gamma), 
\end{equation}
where $F(\gamma) = \int\limits_0^1 J(u) u^3du- \gamma J(0)/2 + 1$. The function $F(\gamma)$ determines also the excitation spectrum of valley current since $\gamma \propto 1/\omega^2$. $F(\gamma)$ calculated numerically from Eq.~(\ref{111}) is shown in Fig.~2. The estimation for $\hbar\omega=1$~eV and  $N_i=10^{12}$~cm$^{-2}$ yields $\gamma \sim 10^{-2}$ and $\chi^{({\rm ee})}$ being two orders of magnitude larger than $\chi^{({\rm vel})}$. 
Thus, for graphene on a substrate with high susceptibility, the mechanism of valley current formation caused by electron-electron scattering dominates. 

To summarize, we have shown that the homogeneous excitation of graphene with a linearly polarized light results in a pure valley current and developed the microscopic theory of this effect.
Pure valley current is accompanied by no net charge current, but leads to accumulation of valley-polarized carriers at edges of the sample. The valley polarization breaks the time inversion symmetry and also implies the local lowering of space symmetry to the $D_{3h}$ group of a single valley, which lacks the space inversion. The space symmetry lowering can be detected by optical means, e.g., by a second harmonic generation of the probe beam. Another possibility to register the valley current is to convert it into an electric current, which can be realized in curved graphene. The curvature of the graphene sheet produces effective out-of-plane magnetic fields directed oppositely for electrons in the valleys $K$ and $K'$.~\cite{Guinea10} Owing to the Lorentz force, the magnetic fields change the directions of partial currents in the valleys giving rise to a measurable net electric current.

We thank E.\,L.\,Ivchenko, D.\,L.\,Shepelyansky, and A.\,D.\,Chepelianskii for stimulating discussions. The work was supported by RFBR, President grant for young scientists, and ``Dynasty'' Foundation -- ICFPM.

\end{document}